\documentclass[manuscript]{aastex}

\shorttitle{ } \shortauthors{Wang et al.}

\begin{document}
\title{Origin of power-law X-ray emission in the Steep power-law state of X-ray Binaries}
\author{Jiancheng Wang and Lihong Yan\altaffilmark{1}}
\affil{Yunnan Observatory /National Astronomical Observatories,
Chinese Academy of Sciences, P.O. Box 110 Kunming, Yunnan Province
650011, P.R. China}
\altaffiltext{1}{Graduate School, Chinese
Academy of Sciences, Beijing, P.R. China }

\begin{abstract}
We present a new scenario of the emissive origin in the Steep Power
Law (SPL) state of X-ray Binaries. The power-law component of X-ray
emission is the synchrotron radiation of relativistic electrons in
highly magnetized compact spots orbiting near the inner stable
circular orbit (ISCO) of black hole and has a hard spectrum that
extends to above MeV bands determined by electron acceleration rate.
These photons are then down-scattered by the surrounding plasma and
form an observed steep spectrum. The relevance of this model with
high-frequency quasi-periodic oscillations (HFQPOs) and extremely
high luminosity of the SPL state is discussed.
\end{abstract}

\keywords{accretion, accretion disks---black hole
physics---radiation mechanisms: nonthermal---X-rays: binaries}

\section{Introduction}
The physical origin of the SPL state generating HFQPOs, extremely
high luminosity and spectra that extend to above MeV, remains one of
the outstanding problems in Black-hole binaries (BHBs). Most models
on the spectra of the SPL state invoke inverse Compton scattering of
seed photons from the disk in a nonthermal corona
\citep{Zdziarski04, Zdziarski05}. The origin of the nonthermal
electrons has led to the models with complicated geometry and
feedback mechanisms. Alternative models involve bulk motion
Comptonization in the context of a converging sub-Keplerian flow
within $50R_g$ of the Black Hole (BH) \citep{Titarchuk02,
Turolla02}. The frequencies of HFQPOs are as fast as the dynamical
ones at the ISCO of Schwarzschild BHs. The models of HFQPOs involve
the resonance oscillation modes occurring in specific radii where
the radiating source orbit has the coordinate frequencies scaling
with a defined ratio in the Kerr metric \citep{Merloni99,
Abramowicz01, Remillard02, Abramowicz03, Kluz04, Sch04, Tk04, Sch05,
Sch06}. In \cite{Sch04} and \cite{Sch05} they have developed a
geodesic hot spot model to explain the X-ray light curves, in which
a collection of "hot spots", small region of excess emission move on
geodesic orbits near the ISCO. The hot spot model is characterized
by the black hole mass and spin, the disk inclination angle, and the
simple properties of the hot spot (such as size, shape and
overbrightness).

Current models of the SPL state are deficient in terms of specifying
the radiation mechanisms that would imprint a given oscillation mode
into the X-ray light curve, such as the radiation component of hot
spots. In fact the HFQPOs are commonly tied to a hard X-ray
component, rather than the thermal component that can be directly
attributed to the accretion disk. It is unlikely that the HFQPOs are
coming from a cool, thermal spot getting upscattered by a hot
corona. From the photon spectra of the SPL state, there appears to
be a hot corona with Compton parameter $y\sim 1$, the lower energy
bands (2-6KeV) have the smaller amplitude modulation. It is
difficult to explain the larger significance of HFQPO detections in
the higher energy bands (6-30KeV) relative to the signal in the
lower energy bands (2-6KeV).

In this paper, we propose a new scenario that the X-rays are the
synchrotron radiation of highly magnetized compact spots orbiting
near the ISCO, and have a hard spectrum that extends to the MeV
bands determined by electron acceleration rate. These photons are
then down-scattered by the surrounding plasma and form an observed
steep spectrum. In the innermost disc region the spot moves like a
test particle in the black hole field and modulates the observed
X-ray flux with HFQPOs through its synchrotron emission. The
coordinate frequencies of the orbital motion with a defined ratio
present a pair of HFQPOs \citep{Sch04, Sch05}. In section 2 we
present the basic relations determining the synchrotron radiation of
a highly magnetized compact spot orbiting near the ISCO and Compton
scattering mechanism of the SPL formation. In section 3 we conclude
a discussion of open question on the SPL state based on new
scenario.

\section{Synchrotron Emission and Compton Scattering Mechanisms}
We assume that a radiation spot with the size $R_b$ in the comoving
frame, has a Doppler factor $\delta$ due to its motion relative to
the observer. In the observer frame, the total energy of the spot is
related to the comoving energy density $\varepsilon$ by $E\simeq
\varepsilon R_b^3 \delta$. If a fraction $f$ of this energy is
radiated during the observed timescale $R_b/(c\delta)$, the observed
luminosity is given by
\begin{equation}
L\simeq 4\varepsilon f R_b^2 c \delta^4,
\end{equation}
where the factor of $\delta^2$ comes from the observed power with
beaming effect. We assume that a fraction $\varepsilon_B$ of the
energy density $\varepsilon$ is magnetic, we then obtain the
magnetic field
\begin{equation}
B\simeq (\frac{8\pi \varepsilon_B L}{f R_b^2 c \delta^4})^{1/2},
\end{equation}
and the ratio of the radiative cooling time $t_{syn}$ to the
dynamical timescale $t_d$,
\begin{equation}
\frac{t_{syn}}{t_d}=\frac{6\pi m_e c}{\sigma_T \gamma B^2}=3.0\times
10^{-3}(\frac{L}{L_{Edd}})^{-1} (\frac{R_b}{R_g})
(\frac{f}{\varepsilon_B}) \gamma^{-1} \delta^4,
\end{equation}
where $L_{Edd}$ is the Eddington luminosity of the black hole, $R_g$
is the Schwarzschild radius and $\gamma$ is the Lorentz factor of
the electron. The condition of efficient synchrotron is that the
dynamical time of electrons $t_{d}$ is larger that the synchrotron
cooling time of electrons $t_{dyn}$. If $\varepsilon_B$ and $f$ are
the same amount and $L/L_{Edd}$ is not small, the synchrotron
radiation of relativistic electrons will become an effective
mechanism of X-ray radiation in a highly magnetized compact spot
with the size $\Delta R\sim R_g$ near the ISCO. We can assess
whether the Doppler factor $\delta$ is large through the velocity
$v_{\phi}$ of the spot orbiting near the ISCO,
\begin{equation}
v_{\phi}=\frac{r \sin i}{2\pi (r^{3/2}\pm a)},
\end{equation}
where $a$ is the spin momentum of black hole, $i$ is the inclination
angle of the spot orbiting plane with respect to the spin axis of
black hole. We then obtain that the maximum velocity observed is
less than ten percent of light speed. Therefore the beaming effect
is not important. In the following discussion, we ignore the beaming
effect. The electrons of the spot orbiting near the ISCO are assumed
to be accelerated by a specific mechanism. The spectrum of the
accelerated electrons determines the energy spectrum of synchrotron
radiation. The most general assumption for the spectrum of
accelerated electrons is the power-law spectrum with an exponential
cutoff at energy $\gamma_0$:
\begin{equation}
N(\gamma)=N_0 \gamma^{-\alpha} exp(-\gamma/\gamma_0).
\end{equation}
The energy cutoff $\gamma_0$ in the spectrum of electrons is
determined by the balance between the electron acceleration and
cooling times. The acceleration time of electrons is usually given
by the following general form:
\begin{equation}
t_{acc}=\eta \frac{\gamma m_e c}{e B_\perp}.
\end{equation}
Hereafter it is assumed that the magnetic field is distributed
isotropically, e.g. $B_{\perp}=\sqrt{2/3}B$. The parameter $\eta\geq
1$ presents the rate of acceleration which could be the
energy-dependent. In different astrophysical environments, $\eta$
remains a rather uncertain model parameter due to highly unknown
acceleration mechanism. The synchrotron cooling time is given by
\begin{equation}
t_{syn}=\frac{6\pi m_e c}{\sigma_T \gamma B^2}.
\end{equation}
Then the energy cutoff $\gamma_0$ is determined by the condition
$t_{acc}=t_{syn}$:
\begin{equation}
\gamma_0=\sqrt{\frac{6\pi e}{\sigma_T}}B^{-1/2} \eta^{-1/2}.
\end{equation}
The high energy cutoff of synchrotron emission is given by
\begin{equation}
\epsilon_0=(\frac{3}{2})\frac{eBh\gamma_0^2}{2\pi m_e
c}=\frac{9}{4\alpha_f}m_e c^2 \eta^{-1}=160\eta^{-1}MeV,
\end{equation}
which depends only on the parameter $\eta$, where $\alpha_f=1/137$
is fine-structure constant. Thus, in the regime of acceleration with
$\eta \leq 100$, the synchrotron radiation emitted by a highly
magnetized compact spot can result in an effective production of
hard X-rays that extend to above MeV bands. The non-relativistic or
relativistic shock acceleration is most widely accepted theory for
the acceleration of nonthermal particles \citep{Blandford87,
Heavens88, Kirk00}. The produced electron spectrum has the spectral
index of $\alpha\simeq 2.5-3.0$, typically found in the optically
thin synchrotron emission of AGNs. The hard X-rays created by
electron synchrotron emission has the photon index of
$\alpha_p\simeq 1.7-2.0$, which is not consistent with the observed
index of SPL state.

In fact the hard X-ray synchrotron photons will be down-scattered by
the surrounding corona and form a softer spectrum where the photon
index increases by unit $\alpha_p \rightarrow \alpha_p +1$. Since
the energy of hard photons is higher than the electron energy in the
corona, the Comptonization is described by the following equation
\citep{Sunyaev80}
\begin{equation}
\frac{1}{Z^2}\frac{d}{dZ}Z^4 N - \beta N=-\beta f(Z)/Z^3,
\end{equation}
where $Z=\frac{h\nu}{m_e c^2}$,
$\beta^{-1}=\frac{3}{\pi^2}(\tau_0+\frac{2}{3})^2$ corresponding to
the mean number of scatterings, $\tau_0$ is the optical depth
through the corona, and $f(Z)$ is the original spectrum of the
photons. The solution of the above equation is as follows:
\begin{equation}
F_{\nu}(Z)=\frac{\beta}{Z} exp(-\beta/Z)\int^{\infty}_{Z} f(Z)
exp(\beta/\zeta) \frac{d\zeta}{\zeta}.
\end{equation}
For a power law spectrum of the hard photons $f(Z)\propto
Z^{-\alpha_p}$ with $Z/\beta\gg 1$ corresponding to large optical
depth ($\tau_0\gg 1$) and high energy photon ($Z\sim 1$), we have
the Compton scattered spectrum $F_{\nu}(Z)\propto Z^{-\alpha-1}$
that can explain the observed spectra with the photon index of
$2.4-3.0$.

For hard X-ray synchrotron photons ($h\nu \gg 1$), the net effect of
Compton scattering process is a transfer of energy from the photon
to the electron. The average fractional energy change per scattering
is large for high energy photons, the spectral shape at higher
energy bands is seriously affected by Compton scattering. It implies
that most HFQPOs appear more significantly in higher energy bands.
The light curves actually have large amplitude fluctuations in the
higher energy bands, as these scattered photons get smoothed out
less in time. Furthermore the higher harmonic modes are successfully
damped in the scattering process, so is the fundamental peak
\citep{Sch06}.
\section{DISCUSSION}
We have proposed a new scenario of the emissive origin in the SPL
state. Our model is based on two standard processes, in which the
X-ray photons with a power law spectrum extending to above MeV bands
is firstly produced by the synchrotron radiation of highly
magnetized compact spots orbiting near the ISCO and then these
photons are down-scattered by the thermal electrons of surrounding
corona to form an observed steep spectrum. The high energy cutoff of
the spectrum is only determined by the electron acceleration rate in
the spot. As in the model, we show that the spot must be small ($R_b
\sim R_g$) and high magnetic to become an efficient synchrotron
source. Since the efficiency and the absolute flux ($\propto R_b^2
B^3$) of synchrotron radiation contributing to X-ray power law
component depend on the luminosity $L/L_{Edd}$. It implies that the
SPL state tends to dominate BHB spectra as the luminosity approaches
the Eddington limit \citep{McC06}.

In the compact spot, the synchrotron emission and Compton scattering
mechanisms appear capable of producing the SPL spectrum. As the hot
spot model proposed by \cite{Sch04} and \cite{Sch05}, the
synchrotron spots with different lifetimes move on geodesic orbits
near the ISCO, they can produce the X-ray light curve with the HFQPO
power spectra. It is why should the HFQPOs appear in the SPL state
and not others. The high energy photons tend to transfer energy into
the electrons in the surrounding plasma to form steeper energy
spectra, measuring the energy spectra of the different QPOs peaks
would prove extremely valuable in understanding the emission and
scattering mechanisms. There are some important messages in the
relationship between the power-law emission and the QPO properties.
In the sources that exhibit the HFQPOs with a 3:2 frequency ratio,
the low frequency QPO appears when the power law flux is very
strong, whereas the high frequency appears when the power law flux
is weaker \citep{Remillard02, Remillard06}. This implies that the
power law photons most likely are coming emissive spots, and are
affected by Compton scattering as these scattered photons from the
spots of outward orbits with low frequency QPO get smoothed out
less. The fact that the HFQPOs are seen most clearly in high energy
bands challenges that the hard X-ray photons are coming from the
thermal photons of accretion disk getting upscattered by a hot
corona. Since the Compton scatterings seriously damp the amplitude
modulation of the light curves in high energy bands.

\acknowledgments

We acknowledge the financial supports from the National Natural
Science Foundation of China 10673028 and 10778702, and the National
Basic Research Program of China (973 Program 2009CB824800).

\end{document}